# Vector Beams for Real-Time Monitoring of Crystalline Symmetry on $\mu$m Length Scale

**YONGBO LV (吕涌波), ZIDU YU (郁子都), HAOTIAN ZHANG (张昊天), LIEJIA QIAN (钱列加), HAO CHU (储灏)***

[1]*School of Physics and Astronomy, Shanghai Jiao Tong University, Shanghai 200240, China*
*[haochusjtu@sjtu.edu.cn]

**Symmetry constrains and determines the physical properties of crystalline solids. Determining crystalline symmetry often requires large-scale diffraction facility and advanced analysis software and therefore cannot be realized with real-time feedback. Alternatively, optical anisotropy has also been exploited for elucidating the rotation, mirror and inversion symmetries of solids on smaller tabletop setups. Second harmonic generation (SHG), in particular, has demonstrated a remarkable sensitivity to subtle changes in these symmetries during phase transitions. To access the full nonlinear susceptibility of a crystal in a cryogenic or vacuum environment, angle-by-angle rotation of the scattering plane of an obliquely incident and outgoing beam in co-ordination with their polarization states is required. This results in a bulky and wobbly setup, significant feedback latency, and a focal spot size > 20 $\mu$m unsuitable for microscopic samples such as 2D materials. In this work, we make use of vector beams to realize a non-rotating oblique-incidence setup, demonstrating a real-time (up to 15 fps) visualization of SHG rotation anisotropy on a sample area of ~ 2×2 μm². Both the feedback latency and focal spot size represent more than an order of magnitude improvement over previous designs based on a Gaussian beam and mechanically-rotating stages. Here, the use of a vector beam greatly simplifies and reduces the cost of nonlinear optical anisotropy measurements, heralding widespread applications of this approach in both research and industrial settings, for example, for strain monitoring and crystal alignment. Our work may also promote broader integration of structured optical field into solid state and condensed matter research in the future.**

Neumann's principle states that the symmetry of a crystal encodes the symmetry of its physical properties. Therefore, resolving the symmetry of a material is often the first step in understanding and exploiting its mechanical, electrical and optical properties for technological applications. Liquid crystal, optical polarizer, nonlinear optical crystal, ferroelectric material, for example, all derive their functionality from certain sets of broken symmetries in their crystalline structures. These symmetries can be investigated using diffraction-based techniques that require large experimental facility and advanced analysis software. Alternatively, they can also be inferred from a material's susceptibility to external fields, for example electric conductivity, magnetic or optical susceptibility [1–3]. However, linear susceptibility does not always guarantee acute sensitivity to the symmetry-broken state (e.g. the nematic state of $BaFe_2As_2$ parent iron superconductor shows only a few percent difference in electric conductivity between the symmetry-broken *a*- and *b*-axis [4]) or the underlying symmetry at all (e.g. the linear response of a centrosymmetric cubic lattice is fully isotropic).

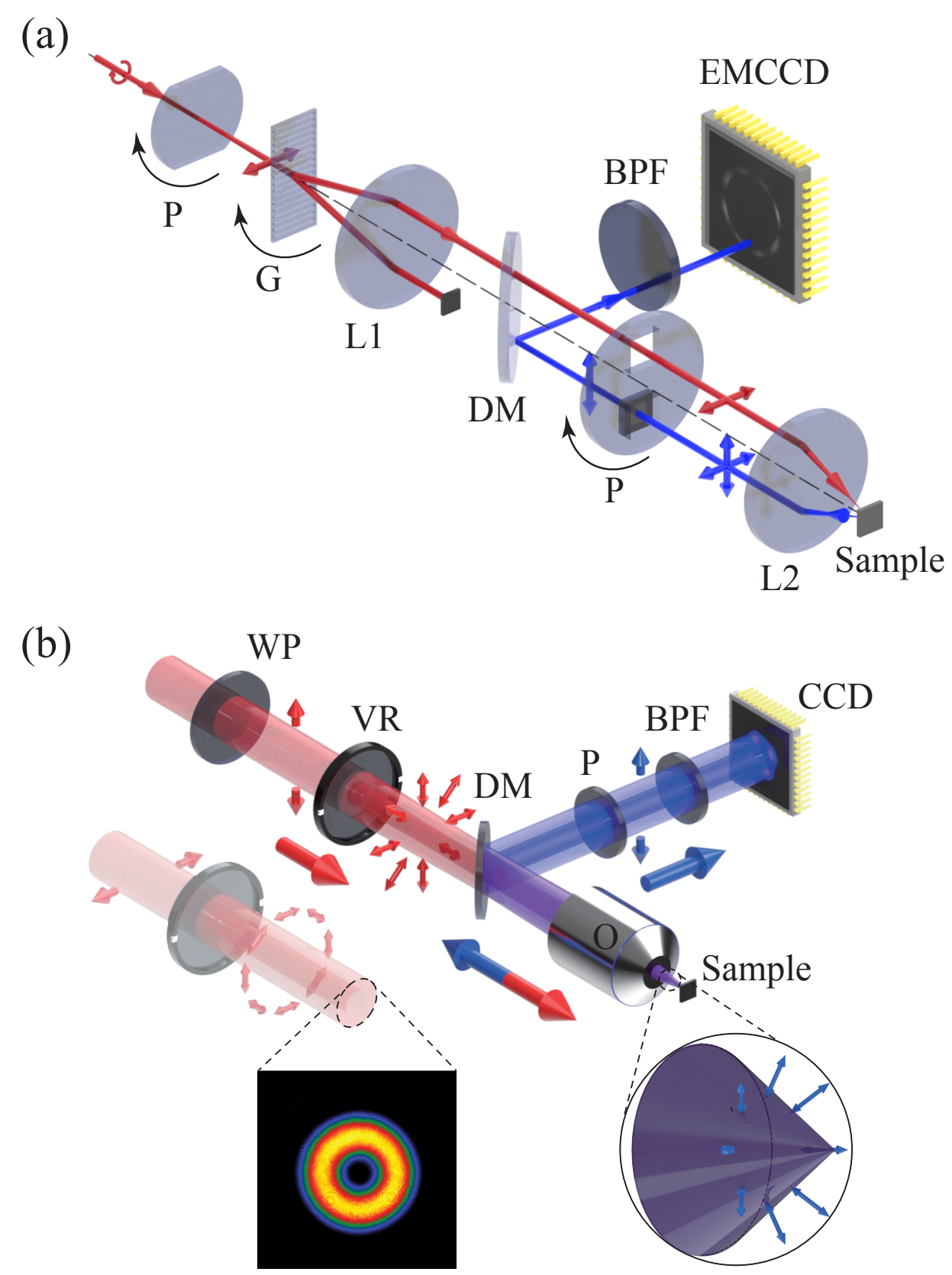

**Fig. 1** Oblique incidence second harmonic generation rotational anisotropy (SHG-RA). **(a)** Previous implementations where a polarizer (P), a diffraction grating (G), a beam block and an analyzer (P) are mechanically rotated together. A CCD is used for recording the SHG-RA pattern. L: lens; DM: dichroic mirror; BPF: bandpass filter; WP: waveplate; O: microscope objective. **(b)** Our design involving no mechanically-moving components. The vortex retarder (VR) transforms a linearly-polarized incoming beam into either a radially- or an azimuthally-polarized beam with a donut-shaped intensity profile. By using a microscope objective with a large numerical aperture, the incident beam effectively acquires an out-of-plane electric field component near the focal point. SHG-RA is obtained from the spatial distribution of the SHG intensity along the azimuth of the beam profile.

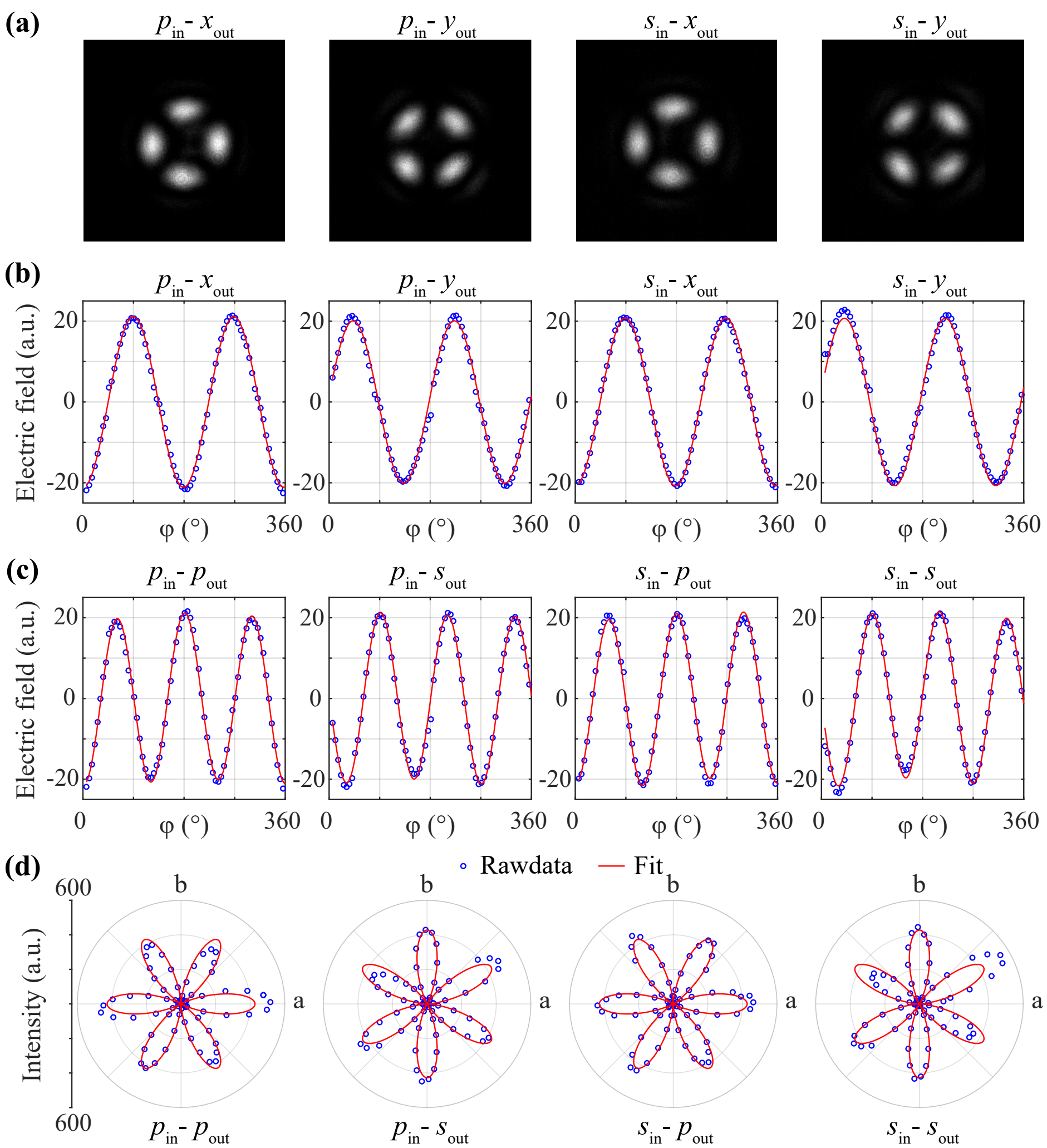

**Fig. 2** SHG-RA from GaAs(111). **(a)** SHG-RA images obtained with {$p_{in}$-$x_{out}$, $p_{in}$-$y_{out}$, $s_{in}$-$x_{out}$, $s_{in}$-$y_{out}$} polarization combinations. **(b)** Electric field profiles of the measured SHG-RA with assumed directions of the electric field. **(c)** Electric field profiles in {$p_{in}$-$p_{out}$, $p_{in}$-$s_{out}$, $s_{in}$-$p_{out}$, $s_{in}$-$s_{out}$} basis calculated from linear combinations of the profiles in (b). **(d)** Reconstructed SHG-RA intensity profiles in {$p_{in}$-$p_{out}$, $p_{in}$-$s_{out}$, $s_{in}$-$p_{out}$, $s_{in}$-$s_{out}$} basis.

In such cases, nonlinear response can be leveraged to provide additional symmetry information and sensitivity. Nonlinear optical response, including second harmonic generation (SHG) and third harmonic generation (THG), has been exploited to visualize the rotation, mirror, inversion symmetries in solids [2,3,5]. In high-temperature superconductors and correlated materials, SHG rotation anisotropy has revealed various symmetry-broken states previously unknown to linear measurements [6–9]. In addition, SHG is also widely used for determining the crystalline orientation of microscopic 2D materials [10–12] as well as for characterizing and monitoring the in-plane strain in various materials [13,14].

In order to capture the full nonlinear susceptibility, for example $\chi^{(2)}_{ijk}(2\omega, \omega, \omega), \{i, j, k\} = \{x, y, z\}$ describing the SHG response, of a solid, experiment requires inputting an obliquely-incident beam and measuring the obliquely-outgoing SHG, so that all $\{x, y, z\}$ components of the susceptibility tensor can be coupled by the incident and outgoing optical field. In some materials, oblique incidence is even necessary for obtaining a non-vanishing SHG response. GaAs, a widely-studied non-centrosymmetric semiconductor with an extraordinary SHG susceptibility, contains only $\chi^{(2)}_{xyz} = \chi^{(2)}_{xzy} = \chi^{(2)}_{yxz} = \chi^{(2)}_{yzx} = \chi^{(2)}_{zxy} = \chi^{(2)}_{zyx}$ as non-zero tensor elements [15]. As a result, GaAs(001) does not generate a finite SHG response to a normally-incident light with electric field lying entirely within the *xy* plane.

In order to implement oblique incidence rotation anisotropy measurements in a cryogenic or vacuum environment (i.e. for studying symmetry-breaking at low temperatures or for protecting the sample against oxidation), often the lab frame (i.e. the optical scattering plane) is rotated to simulate the rotation of the sample. This is realized by rotating a diffraction grating placed on the incoming beam, outputting two mirror-symmetric +1 and -1 order diffracted beams that define the scattering plane (Fig. 1(a)) [16,17]. Polarization-selecting optics on the incoming and outgoing beam paths are rotated in co-ordination with the grating to resolve the optical anisotropy angle-by-angle. This introduces significant latency (typically > 10 s) for experimental feedback. In addition, sophisticated mechanical structures have to be machined to house these optics. A tiny wobble of the rotation-axis may result in the beam walking on the sample surface, making the technique unsuitable for microscopic samples. Additionally, the focal spot size on the sample is fundamentally limited by the use of a convex lens for enabling oblique incidence of the diffracted incident beam, reaching a diameter of > 20 µm in typical implementations [16].

To resolve these issues and to enable real-time visualization of SHG rotation anisotropy on µm length scale, we propose that structured optical fields, in particular vector beams, circumvent these problems and help realize a nonlinear optical anisotropy setup that is extremely simple and low-cost to implement. As shown in Fig. 1(b), a vector beam manifests a spatially-varying polarization profile described by the Poincaré index [18,19]. This provides the spatially-rotating polarization needed for measuring optical anisotropy. Below we will use GaAs as an example to show that vector beam allows the rotational anisotropy of the sample to be accessed instantaneously without the beam rotating or moving on the sample. By combining the vector beam with a microscope objective, oblique incidence measurement can be realized on an area as small as ~ 2 µm in diameter. We show by video that the SHG anisotropy of GaAs(111) can be dynamically visualized in time as the sample rotates (see Visualization 1), unparalleled by any other nonlinear optical anisotropy technique realized so far.

A schematic of our setup is shown in Fig. 1(b). The vector beam is produced from the $TEM_{00}$ Gaussian mode of a Ti:Sapphire amplifier using a commercial vortex retarder. Depending on the incoming beam's linear polarization, the vortex retarder produces either an azimuthally- or radially-polarized vector beam. The singularity of the polarization state in the center of the beam leads to a donut-shaped intensity profile. The vector beam is focused onto the sample using a 50× microscope objective. The beam profile at the focal plane is imaged with a CCD, displaying also a donut shape with a diameter of ~ 2 µm. The large numerical aperture of the objective lens (0.55) implies that the beam is incident onto the sample with a large incidence angle. Therefore, the radially-polarized vector beam acquires a small out-of-plane electric field component and becomes effectively *p*-polarized near the focal plane (note that under high-NA focusing, the *p*-polarization near the focal plane is slightly compromised due to the spatial separation of the *xy*- and *z*-component of electric field. See Supplemental Document Section 2 for simulation and discussion), whereas the azimuthally-polarized vector beam remains *s*-polarized. The nonlinear signal generated is collected by the microscope objective, passing through several wavelength filters and an analyzer that selects its horizontal (*x*-) or vertical (*y*-) component. The anisotropy pattern is finally imaged onto a cooled CCD camera. The resulting SHG-RA pattens in the {$p_{in}$-$x_{out}$, $p_{in}$-$y_{out}$, $s_{in}$-$x_{out}$, $s_{in}$-$y_{out}$} polarization channels can be linearly superposed to obtain the usual {$p_{in}$-$p_{out}$, $p_{in}$-$s_{out}$, $s_{in}$-$p_{out}$, $s_{in}$-$s_{out}$} polarization channels as often reported in literature [7–9,12].

In Fig. 2 we demonstrate the proposed scheme on the surface of a GaAs(111) film, which is expected to generate an SHG response to both *p*-and *s*-polarized incident light. Fig. 2(a) shows the SHG intensity images obtained in the {$p_{in}$-$x_{out}$, $p_{in}$-$y_{out}$, $s_{in}$-$x_{out}$, $s_{in}$-$y_{out}$} basis. Recognizing that the electric field behind the intensity profile switches sign between neighbouring lobes, as indicated by the SHG intensity going through zero between the neighbouring lobes (i.e. two neighbouring SHG lobes are expected to exhibit the same sign in electric field if their intensities do not drop to zero in-between), we can back-calculate the electric field behind the intensity profiles (Fig. 2(b)), which allow us to perform linear combinations to obtain the desired SHG electric field profiles in the {$p_{in}$-$p_{out}$, $p_{in}$-$s_{out}$, $s_{in}$-$p_{out}$, $s_{in}$-$s_{out}$} basis (Fig. 2(c)). The resulting SHG patterns are shown in Fig. 2(d), which agree with the results obtained using the mechanically-rotating setup and with calculation.

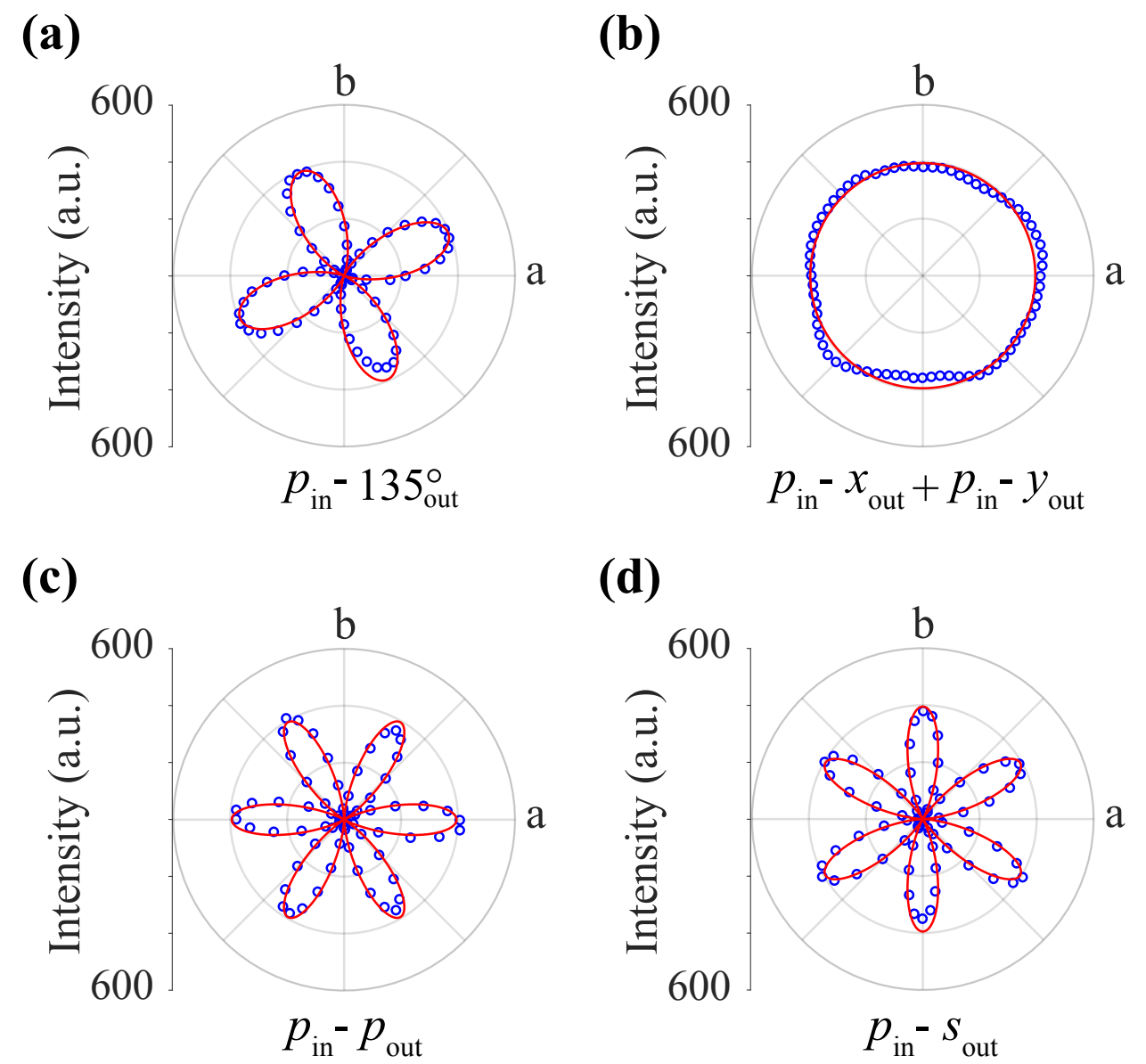

**Fig. 3** Removing the ambiguity in SHG electric field direction by additional $135^{\circ}_{out}$ measurement. **(a)** SHG-RA intensity obtained from GaAs(111) in $p_{in}$-$135^{\circ}_{out}$ (i.e. $p_{in}$-$(E_y - E_x)^2_{out}$) polarization setting. This is contrasted with **(b)** SHG-RA intensity of $p_{in}$-$x_{out}$ + $p_{in}$-$y_{out}$ (i.e. $p_{in}$-$(E_x)^2_{out}$ + $p_{in}$-$(E_y)^2_{out}$). **(c)(d)** SHG-RA intensity in $p_{in}$-$p_{out}$ and $p_{in}$-$s_{out}$ channels are reconstructed from (a)(b) without any assumption of the SHG electric field direction, using the procedure described in the main text and Supplemental Document Section 3.

If the assumption about the sign of the electrical field underlying the neighbouring SHG peaks is dubious, we can also measure additional polarization combinations {$p_{in}$-$135^{\circ}_{out}$, $s_{in}$-$135^{\circ}_{out}$} (in other words {$p_{in}$-$(E_y - E_x)^2_{out}$, $s_{in}$-$(E_y - E_x)^2_{out}$}) to dispose of such ambiguity. Focusing on the *p*-polarized input case: with the additional $p_{in}$ - $135^{\circ}_{out}$ measurement, we let (for a step-by-step derivation, see Eqn. S1–S4 in Supplemental Document Section 3)

$$I_0 = p_{in}x_{out} + p_{in}y_{out},$$
$$I_1 = p_{in}x_{out} - p_{in}y_{out},$$
$$I_2 = I_0 - 2 \cdot p_{in}135^{\circ}_{out}.$$

It is easy to show that

$$p_{in}p_{out} = \frac{1}{2}\{I_0 + I_1\cos(2\varphi) + I_2\sin(2\varphi)\},$$
$$p_{in}s_{out} = \frac{1}{2}\{I_0 - I_1\cos(2\varphi) - I_2\sin(2\varphi)\}.$$

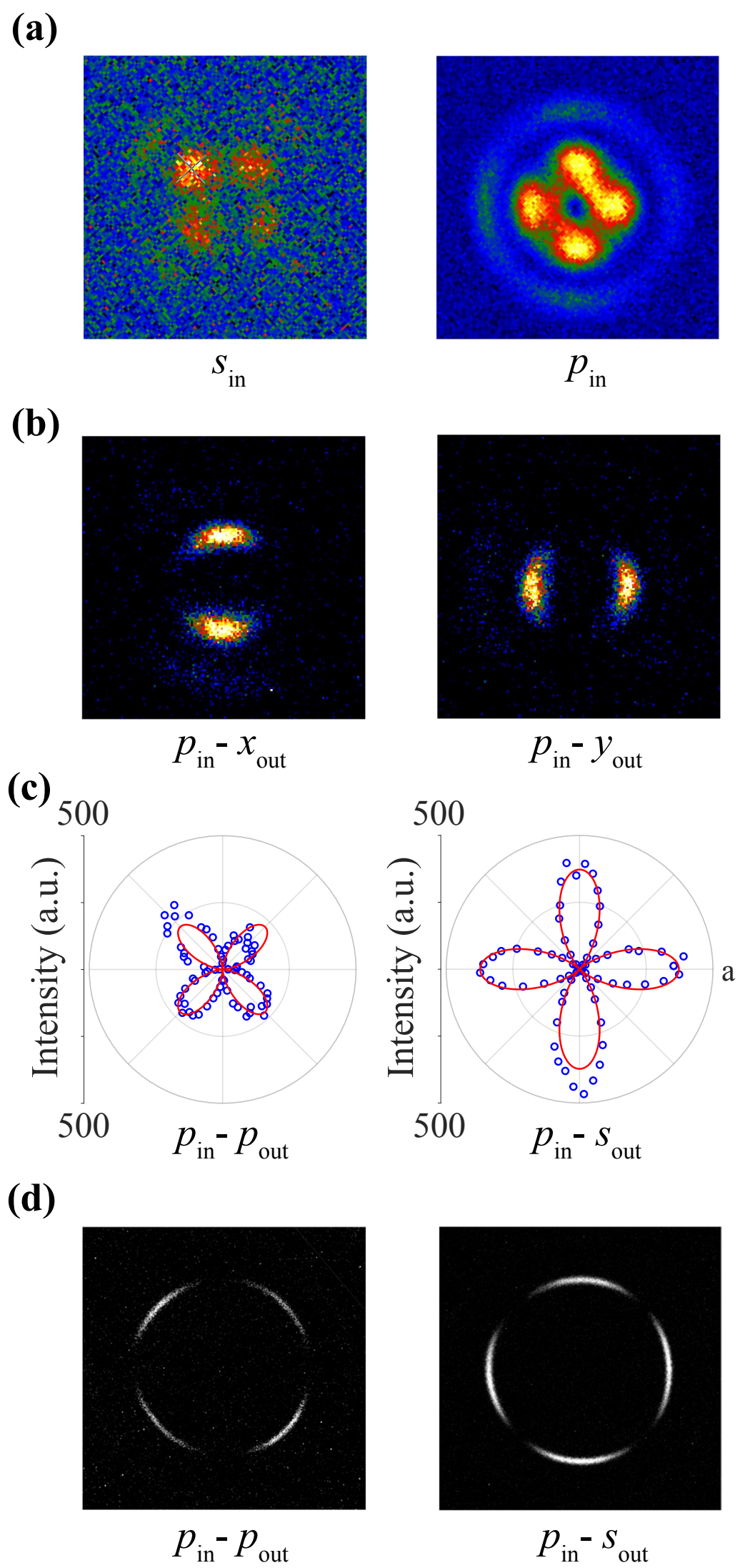

**Fig. 4** SHG-RA from GaAs(001). **(a)** SHG-RA images obtained with $p_{in}$ and $s_{in}$ polarization with no analyzer. SHG by the $s_{in}$ polarization is too weak to be resolved for its *x* and *y* polarization components. **(b)** *x* and *y* component of the SHG generated by $p_{in}$ polarization. **(c)** Reconstructed $\{p_{in}\text{-}p_{out},\ p_{in}\text{-}s_{out}\}$ SHG-RA from GaAs(001). **(d)** $\{p_{in}\text{-}p_{out},\ p_{in}\text{-}s_{out}\}$ SHG-RA from GaAs(001) measured using the previous oblique incidence setup shown in Fig. 1(a).

We demonstrate this idea on GaAs(111). The additional $p_{in}\text{-}135^{\circ}_{out}$ result (Fig. 3(a)) in combination with the results in Fig. 2(a) allows us to obtain $\{p_{in}\text{-}p_{out},\ p_{in}\text{-}s_{out}\}$ without any assumption on the direction of the SHG electric field. The results in Fig. 3(c)(d) confirm the validity of the assumptions made for Fig. 2(b). This method may provide a one-time anchoring of the SHG electric field direction on a new sample before systematic studies of the sample.

To show that we indeed realize a *p*-polarized incident field on the sample surface, we perform measurements on a GaAs(001) surface, which yields a finite SHG response only to a *p*-polarized incident light. Indeed, we observe a negligible SHG response to an azimuthally-polarized vector beam and a strong SHG response to a radially-polarized vector beam (Fig. 4(a)), both as expected. We then perform linear combinations of the results in Fig. 4(b) to obtain the $\{p_{in}\text{-}p_{out},\ p_{in}\text{-}s_{out}\}$ SHG-RA patterns (Fig. 4(c)). A remarkable agreement with the results obtained from a previous SHG-RA setup (Fig. 4(d)) and with calculation is again demonstrated.

Finally, we note that the focal spot size achieved in this work is about an order of magnitude smaller compared to previous oblique incidence setups (> 20 μm). This allows samples with microscopic dimensions, inhomogeneity, or manifesting different symmetry-broken domains, to be better resolved spatially (see Supplemental Document Section 4 for discussion about how to safe check against the possibility of even smaller spatial inhomogeneity < 2 μm in affecting the interpretation of optical anisotropy measured with our technique). Additionally, since no angle-by-angle acquisition is needed, the time for obtaining the SHG pattern can be significantly reduced. For example, using a 200 kHz Yb fiber laser as the excitation source, we can visualize the SHG anisotropy pattern in real time at 15 fps (with 40 ms integration time per frame) on a 2010 Andor Clara CCD (without EM gain), as demonstrated in Visualization 2 (or 2.5 fps with a 1 kHz source as demonstrated in Visualization 3). In cases where the sample does not manifest a strong SHG response, the symmetry-unforbidden THG response may be monitored instead. The remarkably small feedback latency could promise widespread application of our technique in research and industry, where optical anisotropy provides useful feedback for strain, crystal alignment, or even changes in the electronic structure or carrier population. Our work represents one of the first steps in utilizing topological optical fields for solid-state research. We hope that it stimulates a broader integration of structured optical fields including vortex beams into solid-state spectroscopy and condensed matter research in the future [20,21].

**Funding.** National Key Research and Development Program of China (2023YFA1608501, 2024YFA1408701), National Natural Science Foundation of China (12274286), Yangyang Development Fund.

**Disclosures.** The authors declare no conflicts of interest.

**Data Availability.** Data are available upon request to the corresponding author.

## References

1. R. R. Birss, *Symmetry and Magnetism*, 2nd ed. (North-Holland Publishing Co., 1966).
2. M. Fiebig *et al.*, J. Opt. Soc. Am. B **22**, 96 (2005).
3. A. Kirilyuk and T. Rasing, J. Opt. Soc. Am. B **22**, 148 (2005).
4. J.-H. Chu *et al.*, Science **329**, 824 (2010).
5. M. Fiebig *et al.*, Phys. Rev. Lett. **87**, 137202 (2001).
6. J. C. Petersen *et al.*, Nat. Phys. **2**, 605 (2006).
7. D. H. Torchinsky *et al.*, Phys. Rev. Lett. **114**, 096404 (2015).
8. J. W. Harter *et al.*, Science **356**, 295 (2017).
9. A. D. L. Torre *et al.*, Nat. Phys. **17**, 777 (2021).
10. Y. Li *et al.*, Nano Lett. **13**, 3329 (2013).
11. Z. Sun *et al.*, Nature **572**, 497 (2019).
12. H. Chu *et al.*, Phys. Rev. Lett. **124**, 027601 (2020).
13. H. Xing *et al.*, Commun Phys **7**, 382 (2024).
14. S. Sharma *et al.*, Sci. Adv. **12**, eaed4575 (2026).
15. V. V. Pavlov *et al.*, Phys. Rev. Lett. **94**, 157404 (2005).
16. D. H. Torchinsky *et al.*, Rev. of Sci. Instr. **85**, 083102 (2014).
17. J. W. Harter *et al.*, Opt. Lett. **40**, 4671 (2015).
18. D. Pohl *et al.*, Applied Physics Letters **20**, 266 (1972).
19. Y. Mushiake *et al.*, Proc. IEEE **60**, 1107 (1972).
20. A. De Las Heras *et al.*, Optica **9**, 71 (2022).
21. A. García-Cabrera *et al.*, Commun Phys **7**, 28 (2024).

## References

1. R. R. Birss, *Symmetry and Magnetism*, 2nd ed. (North-Holland Publishing Co., 1966).
2. M. Fiebig, V. V. Pavlov, and R. V. Pisarev, Second-harmonic generation as a tool for studying electronic and magnetic structures of crystals: review, Journal of the Optical Society of America B **22**, 96 (2005).
3. A. Kirilyuk and T. Rasing, Magnetization-induced-second-harmonic generation from surfaces and interfaces, J. Opt. Soc. Am. B **22**, 148 (2005).
4. J.-H. Chu, J. G. Analytis, K. De Greve, P. L. McMahon, Z. Islam, Y. Yamamoto, and I. R. Fisher, In-Plane Resistivity Anisotropy in an Underdoped Iron Arsenide Superconductor, Science **329**, 824 (2010).
5. M. Fiebig, D. Fröhlich, Th. Lottermoser, V. V. Pavlov, R. V. Pisarev, and H.-J. Weber, Second Harmonic Generation in the Centrosymmetric Antiferromagnet NiO, Phys. Rev. Lett. **87**, 137202 (2001).
6. J. C. Petersen, M. D. Caswell, J. S. Dodge, I. A. Sergienko, J. He, R. Jin, and D. Mandrus, Nonlinear optical signatures of the tensor order in $Cd_2Re_2O_7$, Nature Phys **2**, 605 (2006).
7. D. H. Torchinsky, H. Chu, L. Zhao, N. B. Perkins, Y. Sizyuk, T. Qi, G. Cao, and D. Hsieh, Structural Distortion-Induced Magnetoelastic Locking in $Sr_2IrO_4$ Revealed through Nonlinear Optical Harmonic Generation, Phys. Rev. Lett. **114**, 096404 (2015).
8. J. W. Harter, Z. Y. Zhao, J.-Q. Yan, D. G. Mandrus, and D. Hsieh, A parity-breaking electronic nematic phase transition in the spin-orbit coupled metal $Cd_2Re_2O_7$, Science **356**, 295 (2017).
9. A. D. L. Torre et al., Mirror symmetry breaking in a model insulating cuprate, Nat. Phys. **17**, 777 (2021).
10. Y. Li, Y. Rao, K. F. Mak, Y. You, S. Wang, C. R. Dean, and T. F. Heinz, Probing Symmetry Properties of Few-Layer $MoS_2$ and h-BN by Optical Second-Harmonic Generation, Nano Lett. **13**, 3329 (2013).
11. Z. Sun et al., Giant nonreciprocal second-harmonic generation from antiferromagnetic bilayer CrI3, Nature **572**, 497 (2019).
12. H. Chu, C. J. Roh, J. O. Island, C. Li, S. Lee, J. Chen, J.-G. Park, A. F. Young, J. S. Lee, and D. Hsieh, Linear Magnetoelectric Phase in Ultrathin $MnPS_3$ Probed by Optical Second Harmonic Generation, Phys. Rev. Lett. **124**, 027601 (2020).
13. H. Xing, J. Liu, Z. Zhao, X. He, and W. Qiu, Quantifying the in-plane strain influence on second harmonic generation of molybdenum disulfide, Commun Phys **7**, 382 (2024).
14. S. Sharma et al., Reconfigurable second-harmonic generation via plasmonic nanoslits counteracting strain-induced suppression in monolayer $MoS_2$, Sci. Adv. **12**, eaed4575 (2026).
15. V. V. Pavlov, A. M. Kalashnikova, R. V. Pisarev, I. Sänger, D. R. Yakovlev, and M. Bayer, Magnetic-Field-Induced Second-Harmonic Generation in Semiconductor GaAs, Phys. Rev. Lett. **94**, 157404 (2005).
16. D. H. Torchinsky, H. Chu, T. Qi, G. Cao, and D. Hsieh, A low temperature nonlinear optical rotational anisotropy spectrometer for the determination of crystallographic and electronic symmetries, Review of Scientific Instruments **85**, 083102 (2014).
17. J. W. Harter, L. Niu, A. J. Woss, and D. Hsieh, High-speed measurement of rotational anisotropy nonlinear optical harmonic generation using position-sensitive detection, Opt. Lett. **40**, 4671 (2015).
18. D. Pohl, Operation of a Ruby Laser in the Purely Transverse Electric Mode TE01, Applied Physics Letters **20**, 266 (1972).
19. Y. Mushiake, K. Matsumura, and N. Nakajima, Generation of radially polarized optical beam mode by laser oscillation, Proc. IEEE **60**, 1107 (1972).
20. A. De Las Heras et al., Extreme-ultraviolet vector-vortex beams from high harmonic generation, Optica **9**, 71 (2022).
21. A. García-Cabrera, R. Boyero-García, Ó. Zurrón-Cifuentes, J. Serrano, J. S. Román, L. Plaja, and C. Hernández-García, Topological high-harmonic spectroscopy, Commun Phys **7**, 28 (2024).

**VECTOR BEAMS FOR REAL-TIME MONITORING OF CRYSTALLINE SYMMETRY ON $\mu$M LENGTH SCALE**

**SUPPLEMENTAL DOCUMENT**

YONGBO LV (吕涌波)[1], ZIDU YU (郁子都)[1], HAOTIAN ZHANG (张昊天)[1], LIEJIA QIAN (钱列加)[1], HAO CHU (储灏)[1*]

[1]School of Physics and Astronomy, Shanghai Jiao Tong University, Shanghai 200240, China

## Section 1 Experimental Setup

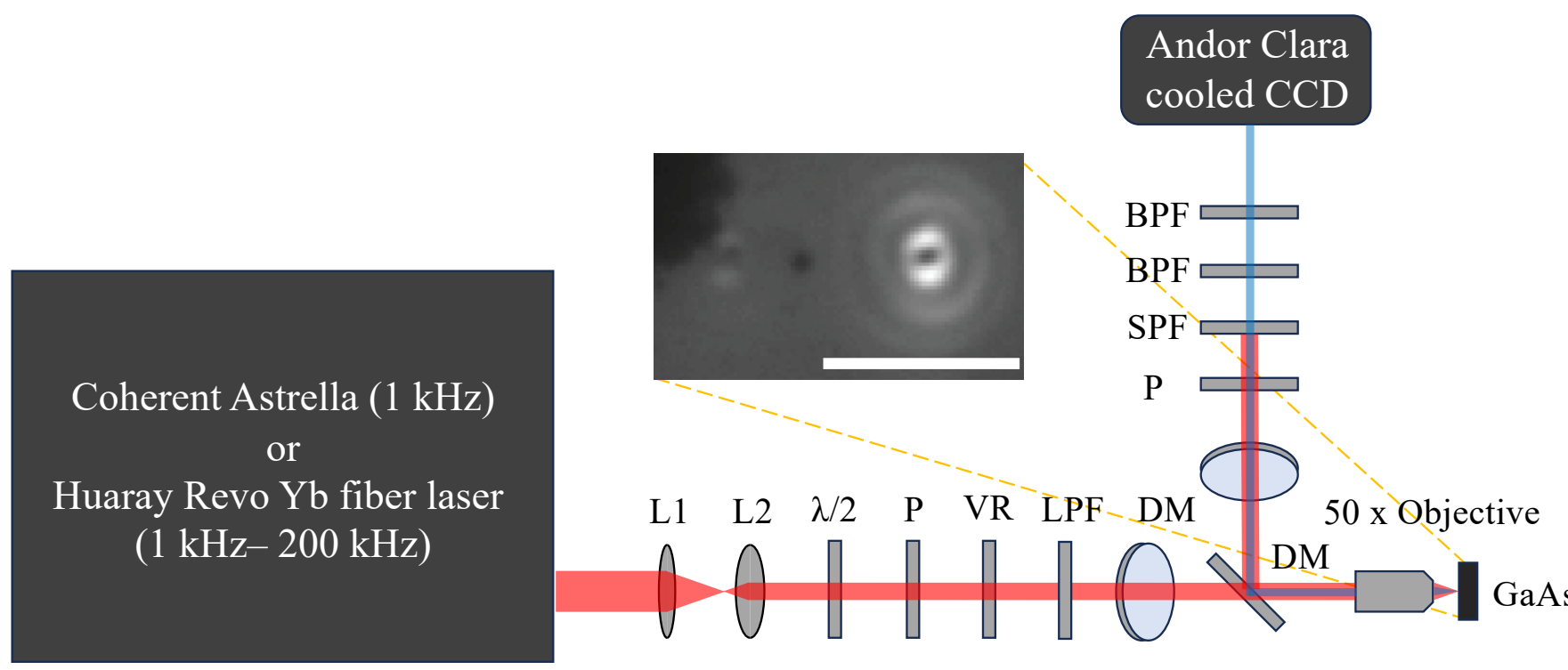

**Figure S1** Layout of the optical beam path. L: lens, $\lambda/2$: half-wave plate, P: polarizer, VR: vortex retarder (LBTEK VR1-795/JCOPTIX LFV1-1030), LPF: longpass filter, DM: dichroic mirror, SPF: shortpass filter, BPF: bandpass filter. Two 45°-tilted DMs are positioned before and after the generation of SHG signal to compensate for the ellipticity introduced by each single DM to the transmitted and reflected linear polarization. An image of the vector beam focus on the GaAs(111) surface is also shown. The scale bar is 10 $\mu$m.

Figure S1 shows a layout of our optical beam path. The results presented in the main text were obtained with an 800 nm 50 fs Coherent Astrella amplifier with 1 kHz repetition rate. The video showing the dynamic feedback of SHG-RA was recorded with a 1030 nm 230 fs Huaray Revo Yb fiber amplifier (tunable between 1 kHz and 200 kHz repetition rate). The SHG-RA patterns are recorded on a 2010 Andor Clara CCD with cooling down to -50 °C but without electron-multiplying capability. We speculate that with more modern EMCCDs with electron-multiplying capability, deeper cooling and faster signal read-out, the dynamic feedback of SHG-RA can be further improved.

We provide two videos comparing the dynamic SHG-RA of GaAs(111) irradiated by the Yb fiber laser set to 100 kHz (Visualization 1) and 1 kHz (Visualization 2) repetition rate. In Visualization 1, the integration time is set to 80 ms and the video refresh rate realized on a ~ 200 pixel × 200 pixel area is 10 fps. In Visualization 2, the integration time is set to 200 ms and the video refresh rate realized on a similar area of CCD is 2.5 fps. If the microscope objective is replaced by a lens so that a bigger sample area can be probed, thereby contributing more SHG intensity, then in theory > 10 fps dynamic visualization may also be possible with a 1 kHz rep rate-laser source.

**Section 2 Simulation of the Electric Field Profile of Radially-Polarized Vector Beam near Focal Point**

In the main text we state that the radially-polarized vector beam acquires an out-of-plane electric field component and becomes effectively *p*-polarized near the focal point due to the large numerical aperture of the microscope objective. However, under strong focusing conditions, the transverse ($E_x$, $E_y$) and longitudinal ($E_z$) components of the electric field may spatially separate near the focal point. Below, we perform numerical simulations of the electric field profile near the focal point to show that a finite *p* polarization is still present near the focal point.

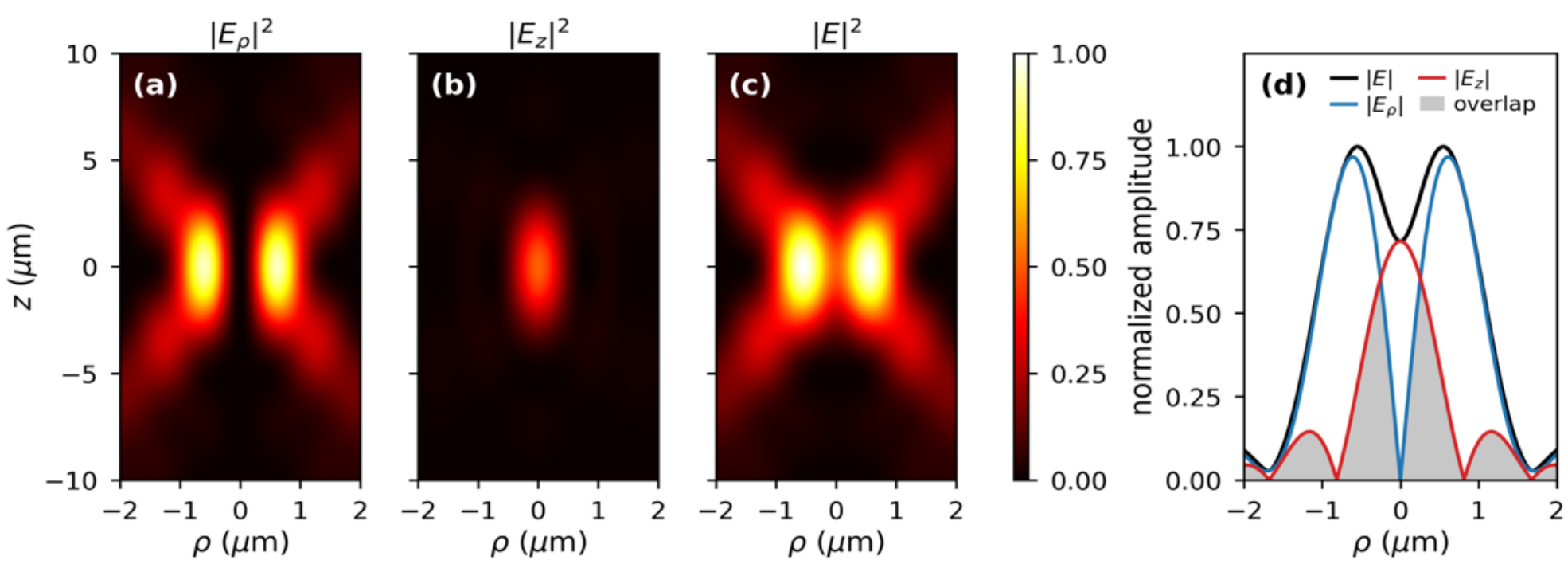

**Figure S2** Simulated electric field profile near the focal point for a radially-polarized vector beam focused by the $50\times$ objective. (a)–(c) $(\rho, z)$ cross-sections of (a) the in-plane component $|E_\rho|^2$, (b) the out-of-plane component $|E_z|^2$, and (c) the total intensity $|E|^2$; $\rho$ is the distance from the beam axis and $z = 0$ is the focal plane. All three panels share the same linear color scale, normalized to the maximum of $|E|^2$. (d) Radial profile of the field amplitude in the focal plane ($z = 0$); the shaded area highlights the overlap region between the in-plane and out-of-plane components, where *p*-polarization is defined.

First, the field arriving at the entrance pupil of the objective is obtained by angular-spectrum propagation of a linearly-polarized Gaussian beam transmitting through the vortex retarder. The focal field is then calculated using Richards–Wolf vectorial diffraction theory [1]. For a radially-polarized vector beam, the focal field is cylindrically symmetric and contains only an in-plane (radial) and an out-of-plane (longitudinal) component:

$$E_\rho(\rho, z) \propto \int_0^\alpha l(\theta)\sqrt{\cos\theta}\,\sin(2\theta)\,J_1(k\rho\sin\theta)\,e^{ikz\cos\theta}\,d\theta,$$

$$E_z(\rho, z) \propto 2i\int_0^\alpha l(\theta)\sqrt{\cos\theta}\,\sin^2\theta\,J_0(k\rho\sin\theta)\,e^{ikz\cos\theta}\,d\theta,$$

where $\theta$ is the convergence angle, $\alpha = \arcsin(\mathrm{NA}/n)$ is the largest angle of incidence on the sample, *k* is the wavevector of the field, $z = 0$ is the focal plane, and $l(\theta)$ is the pupil apodization obtained from the propagated field through the Abbe sine condition ($\rho_{\text{pupil}} = f\sin\theta$).

Simulation parameters: $\lambda = 800$ nm; incoming beam with $1/e^2$ diameter of 4.7 mm; $50\times$ objective with $\mathrm{NA} = 0.55$ ($f = 4.0$ mm, entrance-pupil radius 2.2 mm). Since beam slightly overfills the entrance pupil, the full aperture of the objective is used, corresponding to an angle of incidence of 33° on the sample.

As shown in Fig. S2, the in-plane component $E_\rho$ vanishes on the beam axis and forms the donut profile that is imaged on the CCD, with an outer diameter (FWHM) of $1.9\ \mu m$. The out-of-plane component $E_z$ instead peaks on the beam axis, where $E_\rho$ vanishes. Away from the center, there are regions where $E_z$ and $E_\rho$ are both nonzero. These are the regions where *p*-polarization can be defined, although a significant amount of beam power falls into the region characterized by almost a fully in-plane polarization.

## Section 3 Mathematical Derivation of the SHG-RA Pattern Based on $135^\circ_{\text{out}}$ Measurement

Below, we detail the derivation of the {$p_{\text{in}}$-$p_{\text{out}}$, $p_{\text{in}}$-$s_{\text{out}}$} SHG-RA pattern from the combination of {$p_{\text{in}}$-$x_{\text{out}}$, $p_{\text{in}}$-$y_{\text{out}}$} and $p_{\text{in}}$-$135^\circ_{\text{out}}$ measurement results, without making any assumption for the SHG electric field direction. $E_p$, $E_s$, $E_{\text{x}}$, $E_{\text{y}}$, $E_{45}$ and $E_{135}$ denote the SHG electric field component projected along the radial, azimuthal, x, y, 45° and 135° directions for a given azimuthal angle $\varphi$, as illustrated in Fig. S3.

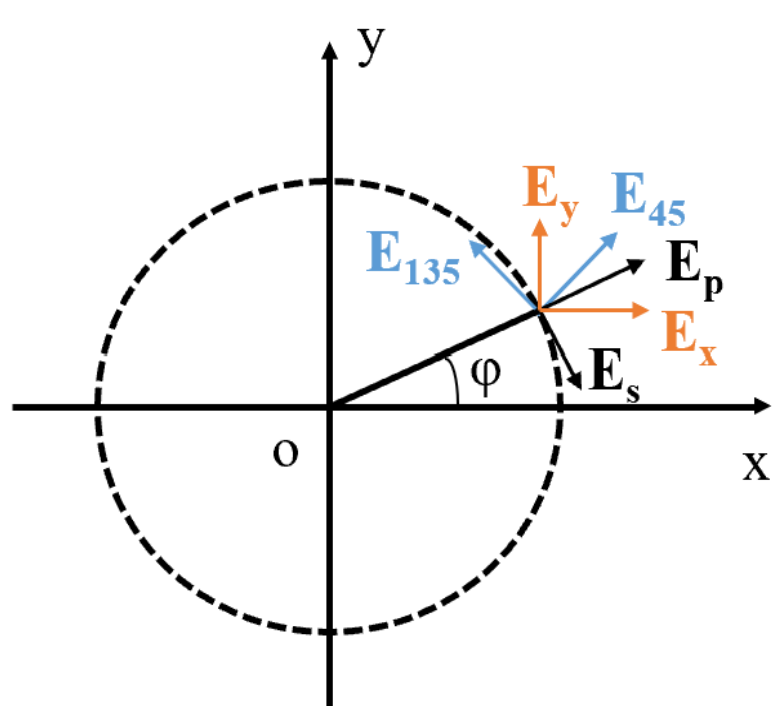

**Figure S3** Definitions of $E_p$, $E_s$, $E_{\text{x}}$, $E_{\text{y}}$, $E_{45}$ and $E_{135}$ used in the following calculation.

First, $E_p$ and $E_s$ are related to $E_{\text{x}}$ and $E_{\text{y}}$ by:

$$\begin{pmatrix} E_p \\ E_s \end{pmatrix} = \begin{pmatrix} \cos\varphi & \sin\varphi \\ \sin\varphi & -\cos\varphi \end{pmatrix} \begin{pmatrix} E_{\text{x}} \\ E_{\text{y}} \end{pmatrix}.$$

We also have

$$I_{\text{x}} = |E_{\text{x}}|^2, \qquad I_{\text{y}} = |E_{\text{y}}|^2, \qquad I_p = |E_p|^2, \qquad I_s = |E_s|^2.$$

Thus

$$I_p = |E_{\text{x}}|^2 \cos^2\varphi + |E_{\text{y}}|^2 \sin^2\varphi + (E_{\text{x}}E_{\text{y}} + E_{\text{y}}E_{\text{x}})\sin\varphi\cos\varphi,$$

$$I_s = |E_{\text{x}}|^2 \sin^2\varphi + |E_{\text{y}}|^2 \cos^2\varphi - (E_{\text{x}}E_{\text{y}} + E_{\text{y}}E_{\text{x}})\sin\varphi\cos\varphi.$$

Setting $E_x E_y = E_y E_x = Q$, we have

$$I_p = I_x \cos^2\varphi + I_y \sin^2\varphi + 2Q\sin\varphi\cos\varphi,$$
$$I_s = I_x \sin^2\varphi + I_y \cos^2\varphi - 2Q\sin\varphi\cos\varphi.$$

Thus

$$I_p + I_s = I_x(\cos^2\varphi + \sin^2\varphi) + I_y(\cos^2\varphi + \sin^2\varphi) = I_x + I_y,$$
$$I_p - I_s = I_x(\cos^2\varphi - \sin^2\varphi) - I_y(\cos^2\varphi - \sin^2\varphi) + 4Q\sin\varphi\cos\varphi$$
$$= (I_x - I_y)\cos 2\varphi + 2Q\sin 2\varphi.$$

Let $I_0 = I_x + I_y$ and $I_1 = I_x - I_y$, we have

$$I_0 = I_x + I_y = I_p + I_s \qquad \text{Eqn. (S1)},$$
$$I_p - I_s = I_1\cos 2\varphi + 2Q\sin 2\varphi \qquad \text{Eqn. (S2)}.$$

Similarly,

$$\begin{pmatrix} E_{45} \\ E_{135} \end{pmatrix} = \begin{pmatrix} \frac{1}{\sqrt{2}} & \frac{1}{\sqrt{2}} \\ -\frac{1}{\sqrt{2}} & \frac{1}{\sqrt{2}} \end{pmatrix} \begin{pmatrix} E_x \\ E_y \end{pmatrix},$$

$$E_{135} = \frac{E_y - E_x}{\sqrt{2}}, \qquad E_{45} = \frac{E_y + E_x}{\sqrt{2}}.$$

Therefore,

$$I_{135} = |E_{135}|^2 = \frac{|E_y|^2 + |E_x|^2 - (E_y E_x + E_x E_y)}{2} = \frac{I_y + I_x - 2Q}{2},$$
$$I_{45} = |E_{45}|^2 = \frac{|E_y|^2 + |E_x|^2 + (E_y E_x + E_x E_y)}{2} = \frac{I_y + I_x + 2Q}{2}.$$

So,

$$I_{45} + I_{135} = I_x + I_y = I_0,$$
$$I_{45} - I_{135} = 2Q,$$
$$Q = \frac{I_0 - 2I_{135}}{2}.$$

Let $I_2 = I_0 - 2I_{135} = 2Q$, from Eqn. (S1) and Eqn. (S2) we have

$$I_p + I_s = I_0 \qquad \text{Eqn. (S3)},$$
$$I_p - I_s = I_1\cos 2\varphi + I_2\sin 2\varphi \qquad \text{Eqn. (S4)}.$$

Summarizing all, we defined

$$I_0 = I_x + I_y,$$
$$I_1 = I_x - I_y,$$
$$I_2 = I_0 - 2I_{135}.$$

Solving Eqn. (S3) and (S4) gives

$$I_p = \frac{1}{2}(I_0 + I_1\cos 2\varphi + I_2\sin 2\varphi),$$

$$I_s = \frac{1}{2}(I_0 - I_1\cos 2\varphi - I_2\sin 2\varphi).$$

**Section 4 Assessing Spatial Inhomogeneity Beyond the Focal Spot Size**

In our implementation of SHG-RA using vector beams, the nonlinear optical signals at different azimuthal angles are collected from different areas of the sample, despite all spots falling within a radius of ~ 1 $\mu$m from the center of the beam. If the sample manifests spatial inhomogeneity smaller than this length scale, for example due to nanoscopic symmetry-broken domains, poly-crystalline domains, or fast-varying strain distribution, it might be of concern whether the symmetry information obtained this way is meaningful or not. For example, if the SHG-RA patten indicates symmetry-breaking along x-direction (e.g. higher SHG intensity in x-direction than y-direction), does it imply a uniform lattice distortion/strain along x-direction with a length scale ≥ 2 $\mu$m, or is it because there is an even finer spatial distribution of lattice distortion/strain along the annulus of the vector beam, whose variation gives rise to the distorted SHG-RA?

To address this concern, we propose the following one-time measurement which can help elucidate whether the sample is homogeneous along the annulus of the vector beam's focus spot. Briefly speaking, we obtain the local SHG-RA associated with each azimuthal angle of the vector beam as we rotate the incoming beam's linear polarization along with the analyzer after the sample. Then, we compare the local SHG-RA from opposite sides of the annulus separated by 180°. The electric fields falling on these two spots are mirror-symmetric of each other and should lead to similar SHG-RA patterns if the sample is homogeneous across these two spots. Pairs of such 180°-separated spots can be checked in sequence to infer the overall homogeneity of the sample along the annulus.

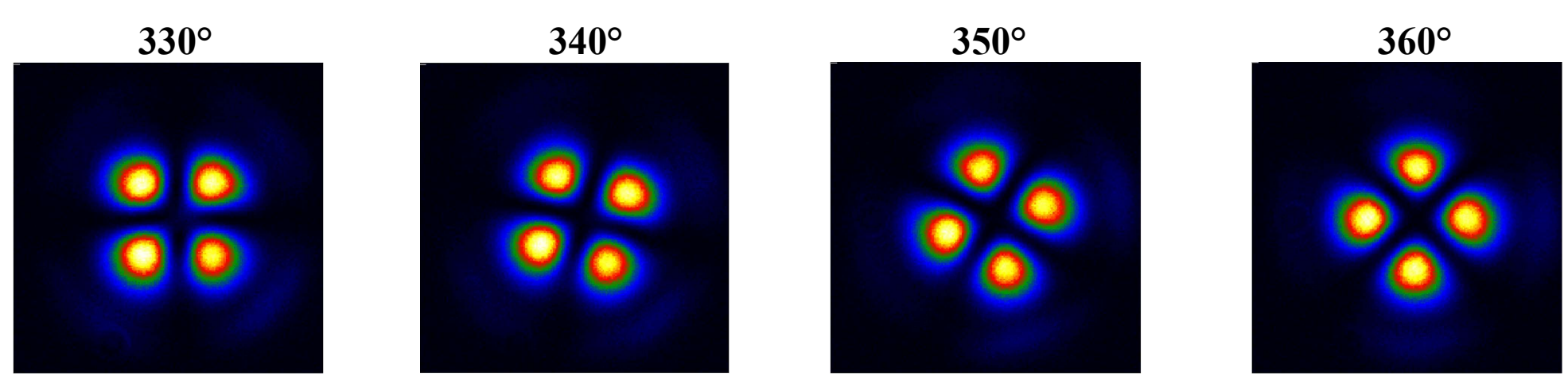

**Figure S4** Examples of SHG-RA images of GaAs(111) recorded in the parallel polarization channel ($E_{2\omega}$ // $E_{\omega}$) for different angles {330°, 340°, 350°, 360°} of the incoming linear polarization (i.e. waveplate angle × 2).

To demonstrate this idea, we rotate the half-wave plate in front of the vortex retarder at 5° steps from 0° to 180°, completing a full 360° rotation of the incoming linear polarization before the vortex retarder. As a result, the incident field on the sample continuously varies between radially-polarized and azimuthally-polarized. The in-plane component of the electric field on the sample continuously varies in direction along a circle, mimicking the acquisition of SHG-RA (although for certain angles of the waveplate the incident electric field also has a finite out-of-plane component). The analyzer is simultaneously rotated at 10° steps from 0° to 360° to project out the parallel-polarized component of SHG. For each angle of the waveplate and the analyzer, the SHG-RA image is recorded, as illustrated in Fig. S4 for a few such angles.

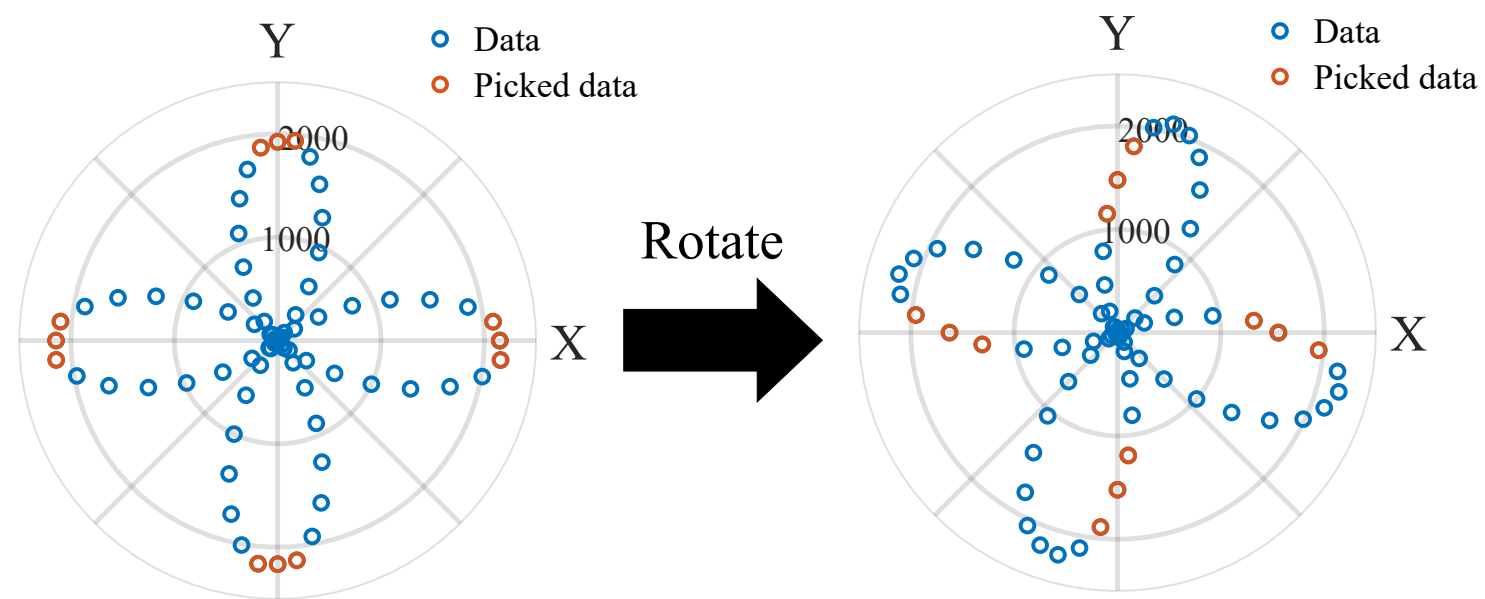

**Figure S5** Selecting pairs of opposite azimuthal angles {0°, 180°} and {90°, 270°} (highlighted in brown) on the SHG-RA data for integrating (± 5° width) the SHG intensity generated from local areas of the sample falling under those angles of the vector beam annulus. The incoming beam's linear polarization is then rotated and the integration is repeated for the same azimuthal angles.

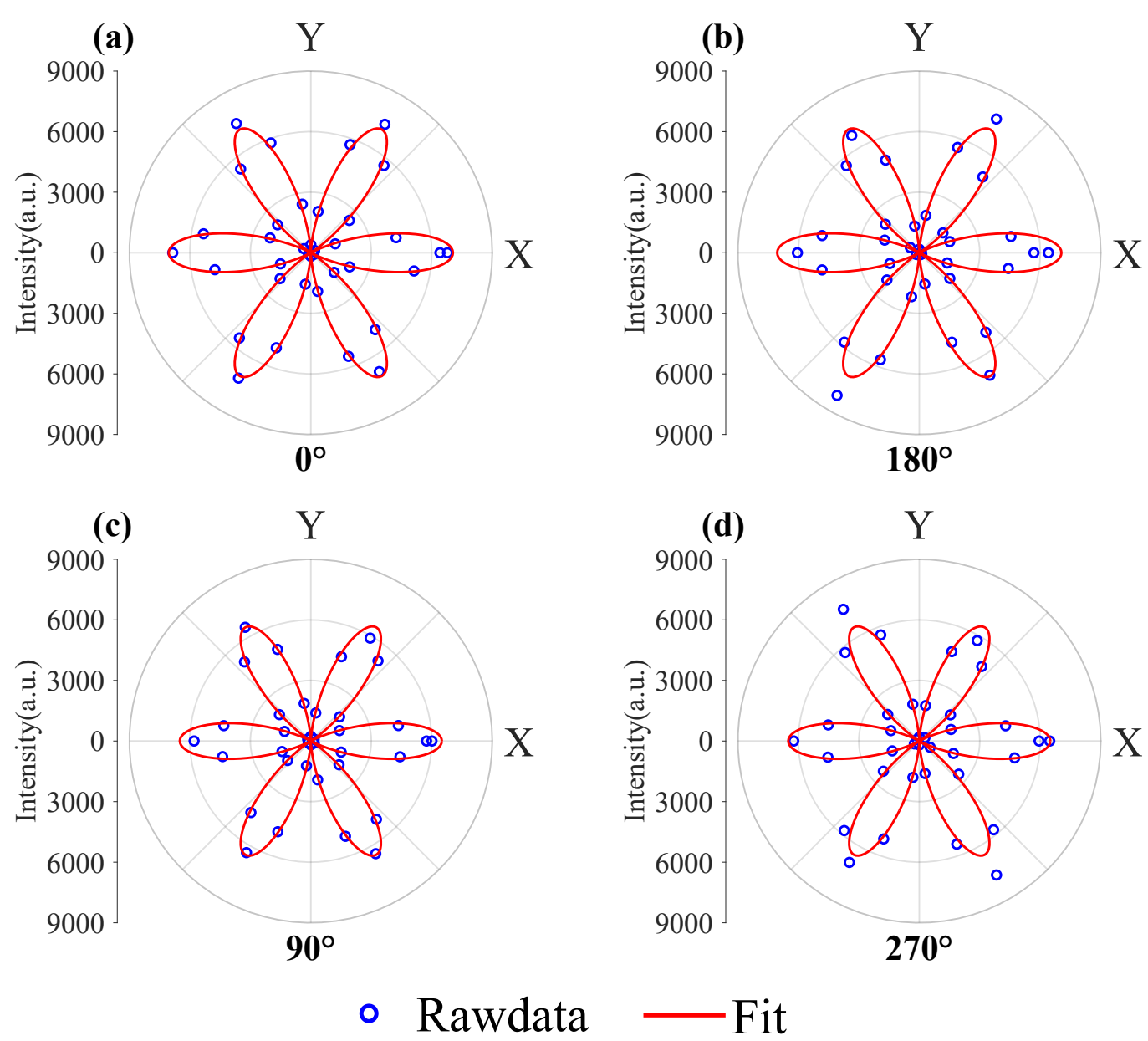

**Figure S6** Local or quasi- SHG-RA patterns obtained from two sets of opposite azimuthal angles {0°, 180°} and {90°, 270°}, based on the raw SHG-RA data shown/illustrated in Fig. S4 and S5.

On each of these recorded SHG-RA images, we can integrate the SHG intensity along certain azimuthal angles, as illustrated in Fig. S5. By plotting each azimuthal angle's SHG intensity as a function of the incoming polarization/analyzer angle, we obtain the local SHG-RA or 'quasi-SHG-RA' (i.e. 'quasi' in the sense that the incoming polarization varies between *p* and *s* instead of maintaining a constant *p* or *s* state) originating from a local area of the sample falling under the vector beam annulus. The results are shown in Fig. S6. On our GaAs(111) sample, we find remarkable similarity for the local SHG-RA from pairs of opposite angles at {0°, 180°} and {90°, 270°}, indicating good homogeneity of the sample across these spots. On any new samples, this check can be performed prior to systematic measurements to determine whether the sample is homogeneous along the vector beam's annulus.

## References

1. K. S. Youngworth and T. G Brown, "Focusing of high numerical aperture cylindrical-vector beams," Opt. Express 7, 77-87 (2000)